\documentclass[twocolumn]{aastex631}
\journalinfo{Accepted for publication in The Astrophysical Journal}
\usepackage{graphicx} 
\usepackage{natbib}

\begin{document}

\title{Mapping the Active Galactic Nucleus Effects on the Stellar and Gas Properties of NGC 5806}
\author{Sophie L. Robbins}
\affiliation{University of Maryland, Department of Astronomy}
\affiliation{University of California Los Angeles, Department of Physics and Astronomy}

\author{Sandra I. Raimundo}
\affiliation{Physics and Astronomy, University of Southampton}
\affiliation{DARK, Niels Bohr Institute, University of Copenhagen}

\author{Matthew A. Malkan}
\affiliation{University of California Los Angeles, Department of Physics and Astronomy}

\begin{abstract}
    It is commonly accepted that active galactic nuclei (AGN) have a strong impact upon the evolution of their host galaxies, but the processes by which they do so are not fully understood. We aim to further the understanding of AGN feeding and feedback by examining an active galaxy using spatially-resolved spectroscopy. We analyze integral field spectroscopy of the active galaxy NGC 5806, obtained using the Very Large Telescope (VLT) Multi-Unit Spectroscopic Explorer (MUSE). We map the dynamics of gas and stars, as well as gas optical emission line fluxes throughout the central 8 x 8 kpc$^2$ of the galaxy. We use emission line ratios to map gas metallicity and identify regions of gas excitation dominated by AGN/shocks or star formation. We also determine the average stellar population age and metallicity, and model the rotation and dynamics of the galaxy. We find that NGC 5806 has a star-forming circumnuclear ring, with a projected radius of $\sim$400 pc. The dynamics of this galaxy are driven by a large-scale bar, which transports gas from the spiral arm to the central ring and potentially fuels the AGN. We also observe AGN-dominated gas excitation up to 3.3 kpc away from the center of the galaxy, showing the extended AGN effect on the gas in the central regions of the galaxy.
\end{abstract}

\section{Introduction}
\label{sec:Introduction}
Active galactic nuclei (AGN) are suspected to influence the formation and evolution of galaxies. Close correlations between supermassive black holes (SMBHs) and the properties of their host galaxies have been established \citep{FerrareseMerritt2000,Gebhardt2000,MarconiHunt2003,HaringRix2004,Bennert2021}, suggesting a co-evolution between SMBHs and their galaxies, presumably via gas accretion and star formation regulation by the AGN \citep{Fabian2012}. The transportation of fuel to AGNs, and feedback from them, have been used in cosmological modelling to explain the evolution of galaxies into what we see in the present universe \citep{Springel2005,Somerville2008,Byrne2023}. However, the means by which AGN affect and are affected by their host galaxies are still not understood. 

Much research has been conducted on specific possible mechanisms of AGN feedback.  Winds and radiation pressure from the AGN create outflows, which push gas and dust away from the center of the galaxy \citep{Fabian2012}. These outflows have been observed in multiple active galaxies \citep{Riffel2009,StorchiBergmann2010,Raimundo2017,daSilva2020,Costa-Souza2024}, and often move along a ionized bicone centered on the AGN \citep{StorchiBergmann2009,Venturi2018}. By blowing gas and dust away from the nucleus and potentially out of the galaxy, these outflows are thought to quench star formation and eventually AGN activity \citep{Somerville2008, Fabian2012}. However, the study of this feedback is complicated by the geometry of AGN and the different wavelengths of light in which different effects can be observed, so the processes are still under debate.

AGN fueling has been similarly studied, and is similarly not yet fully understood. The inflow of gas along large-scale galactic non-axisymmetric potentials, such as galactic bars, has been identified as a major factor of AGN fueling. Theoretical studies have found that galactic bars move gas towards the center of their galaxies \citep{Sormani2015}. This bar-driven gas inflow serves an important role in removing angular momentum from gas, moving it towards the nucleus, and creating a gas reservoir to feed the AGN \citep{Silva-Lima2022}. However, as \citet{Hunt1999}, \citet{Wada2004} and \citet{Silva-Lima2022} have found, a galactic bar by itself is not sufficient to drive gas all the way to the AGN, and there are many barred galaxies without AGNs. There must be additional factors also at work in feeding the AGN, particularly in the central $\leq$ 100 pc of the galaxy. Several such mechanisms which have been explored in previous studies include galaxy interactions and mergers, gravity-driven turbulence in the gas, dynamical friction between gas clumps and stars, and spiral density waves \citep{Wada2004}.

One feature that has been found in many active galaxies is a star-forming circumnuclear ring, often with a radius around 200-500 pc \citep{Genzel1998,StorchiBergmann2007,Dumas2007,Riffel2009, Hicks2013,Costa-Souza2024}. This ring has been theorized to be both a consequence of large-scale gas inflows and a potential source of small-scale AGN feeding; studies have shown that these rings are formed by inward-flowing gas evolving into a ring due to an inner Lindblad resonance, and that clumps from the ring or the entire ring can collapse inward and feed the AGN \citep{Fukuda2000,Wada2004}. Considering how many active galaxies have been found to have this star-forming circumnuclear ring, it is likely that the ring is connected to the AGN in one of three ways: the central ring is either a consequence of the AGN, a consequence of the conditions which trigger an AGN, or a factor itself contributing to triggering the AGN. 

In order to determine the interplay between an AGN and its host galaxy, it is important to determine how the gas emission and dynamics of gas and stars vary in different regions of the galaxy. For this reason, integral field spectroscopy has become an important tool for the analysis of active galaxies. Studies such as the MUSE Atlas of Disks \citep{Erroz-Ferrer2019}, the program from which the data for this paper were obtained, have used spatially-resolved spectra in order to distinguish the contributions of AGN from the effects of star formation, diffuse ionized gas, and other phenomena. We aim to build off of this research and take it a step further, by focusing on one galaxy and thoroughly examining its spatially-resolved properties.

This paper is organized as follows. In section \ref{ObjectSection}, we describe the galaxy examined in this study and the data we use to observe it, and define regions within the galaxy. Section \ref{SectionDataAnalysis} describes the analysis performed on the data in order to measure emission line fluxes and dynamics, and to model the large-scale dynamics of the galaxy. We present the results in section \ref{SectionResults}, discussion in section \ref{SectionDisc}, and conclusions in section \ref{SectionConclusion}.

\section{Object and Data}
\label{ObjectSection}

\subsection{Object}
NGC 5806 is a nearby SAB(b) galaxy at z=0.00449, located at RA = 15:00:00.47 (J2000) and Dec = 01:53:28.3 (J2000). Throughout this paper, we adopt the cosmology-corrected scale 110 pc/arcsec, from the NASA/IPAC Extragalactic Database \citet{NED}. NGC 5806 has an inclination \textit{i} = 58\textdegree and position angle PA = 166\textdegree\ \citep{CLB2023}. The total stellar mass of this galaxy has been measured at log$_{10}(M_{*}/M_{\odot})=10.59$  \citep{Salo2017}, and the diameter is 189.7 arcseconds \citep{NED}.

This galaxy has been classified as a Seyfert 2 galaxy \citep{VeronCettyVeron}, and several studies have observed gas with mixed ionization mechanisms, including AGN emissions and star formation \citep{Carollo2002,Seth2008,Westoby2012,Erroz-Ferrer2019}. Multiple studies have found that this galaxy has a large-scale bar, and a circumnuclear star-forming ring \citep{Carollo2002,Dumas2007,Erroz-Ferrer2019,CLB2023}.

\begin{figure}[h]
\centering
  \includegraphics[width=.8\linewidth]{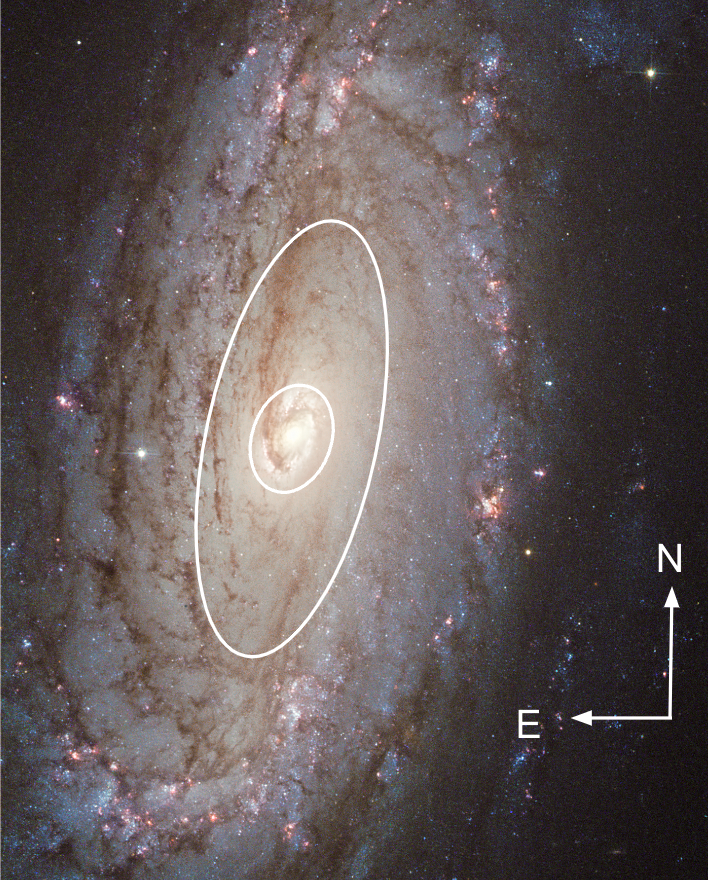}
\caption{Composite Hubble image of NGC 5806, with field of view 1.2 by 1.5 arcminutes. This image combines exposures from the optical filters B (435 nm), V (555 nm), and H$\alpha$ (658 nm), and infrared filter I (814 nm). In this image, one can see supernova SN 2004dg as a yellow spot in the top right corner. This image clearly shows the dark dust lanes in this galaxy, including a dust lane that stretches towards the center of the galaxy and cuts into the central ring. Two white ellipses are superimposed: the inner ellipse outlines the bright nucleus, which is the same central region and central ring seen in our MUSE data. The outer ellipse illustrates the approximate location of the galactic bar in this galaxy. Credit: NASA/ESA. Acknowledgement: Andre van der Hoeven.}

\label{HubbleCompositeImage}
\end{figure}

The composite Hubble image in Figure \ref{HubbleCompositeImage} shows NGC 5806 as seen in the optical and infrared (IR) range. The galaxy has a bright center, a central ring, and a large spiral arm with bright clumps. This image also displays several dust lanes, including one which obscures part of the circumnuclear ring. We see more dust obscuration on the left side of this image than the right; therefore, the left side of this image is the near side of the galaxy, and the right side is the far side. 

This galaxy was previously targeted for a supernova SN 2004dg in its outer reaches \citep{Vagnozzi2004,EliasDeLaRosa2004,Harutyunyan2008}. This supernova can be seen in the top right corner of the Hubble image in Figure \ref{HubbleCompositeImage}. However, this supernova, and any effects it may have had, fall outside the scope of our MUSE data, and thus will not be discussed in this paper. 

\subsection{Data}
For this paper, we analyze spatially resolved integral field unit (IFU) spectra, obtained by the Multi-Unit Spectroscopic Explorer (MUSE) on the Very Large Telescope (VLT) \citep{Bacon2010}. These data were originally collected as part of the program "The MUSE Atlas of Disks" (097.B-0165; PI: C. M. Carollo). The image spans a spectral range 475-935.1 nm with 1.25 $\AA$ sampling. It has sky coverage 2.2 arcmin$^2$ with $\sim$0.6" spatial resolution and 0.2" pixel scale. The galaxy is 22.80 $\pm$ 1.61 Mpc away \citep{NED},, so the data have a physical scale of 22 pc per pixel. These data came from a single exposure, taken with 3600 seconds of integration, taken on April 7, 2016. We used the Phase 3 datacube that was processed by ESO using the standard pipeline and made available on the ESO archive (http://archive.eso.org). For more information on the data reduction, see \citet{DenBrok2020}.

\subsection{Definition of Regions}
\begin{figure}[h]
\centering
  \includegraphics[width=.8\linewidth]{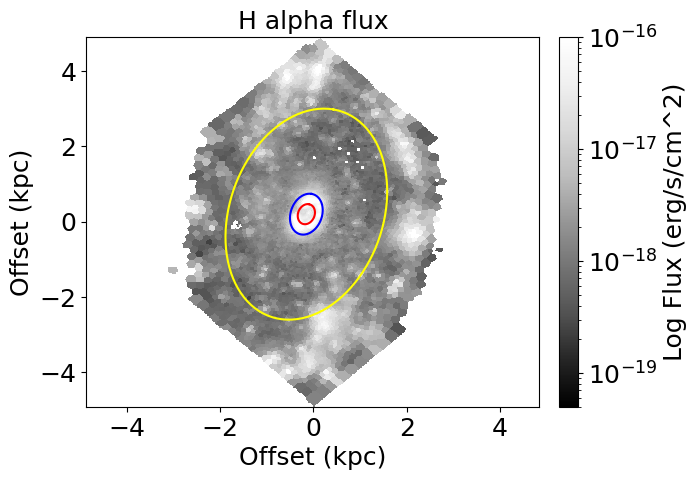}
\caption{A map of H$\alpha$ emission, with ellipses superimposed to define the regions of the image. In the center, within the red ellipse, is the central region or nucleus. Between the red and blue ellipses is the central ring. The area between the blue ellipse and the yellow ellipse is defined as the intermediate region, and the area outside of the yellow ellipse is the spiral arm. These are a rough definition, meant to aid the reader in understanding the regions referred to for the rest of this paper.}
\label{RegionsFig}
\end{figure}

In order to clarify terms for the rest of this paper, we define 4 major regions in our image of this galaxy. These regions are illustrated in figure \ref{RegionsFig}. The "central region" refers to the center of the galaxy, within a projected radius of $\sim$200 pc. Around this, we have the "central ring," with a projected semi-major axis of $\sim$400 pc and an annular thickness of $\sim$200 pc. At the edge of the image, at a projected distance of $\sim$3000 pc from the center, is a spiral arm. The region in between the spiral arm and the central ring will be referred to as the "intermediate region." 

\section{Data Analysis}
\label{SectionDataAnalysis}
We analyzed the datacube to extract the spatially-resolved stellar kinematics, gas dynamics, and dominant ionization mechanisms throughout the galaxy. In order to do so, we binned the data, used pPXF to model the spectra of each bin, and used DiskFit to model the disk rotation. Each of these steps will be discussed below.\par

\subsection{Voronoi Binning}
First, we spatially binned the data, in order to increase the signal to noise ratio (SNR). We used the Voronoi binning technique as implemented in \citet{Cappellari2003} to bin spaxels, so that each bin had a SNR $\geq$ 50. We took into account the flux uncertainties in each spaxel so that spaxels with high uncertainties or non-physical values had a lower weighting when binning the datacube. As input to the Voronoi binning we used the data and error spectra from the data cube in each spaxel. We evaluated the mean SNR in the rest range 6650-6700$\AA$, where there should be no significant emission lines, in order to evaluate the SNR of the continuum which was then used for the Voronoi binning. This produced binned spaxels with SNR $\geq$ 50. Our final binned data cube contains the average spectrum within each bin. 

\subsection{Fitting Models to Spatially-Resolved Spectra}
\label{PPXFsection}
We used the Python package pPXF \citep{Cappellari2023} to fit a model consisting of velocity-broadened stellar templates and emission lines to the spectrum in each bin of the data cube. For this process, we analyzed the spectrum in the rest range 4800-8800$\AA$, as the spectrum at wavelengths larger than 8800$\AA$ was contaminated with large residuals from sky subtraction.  

For stellar templates, we used the XShooter release 2 library of stellar spectra  \citep{Gonneau2020}, which can be found at http://xsl.u-strasbg.fr/index.html. These templates were chosen due to their narrow full-width half-maximum (FWHM=0.598 $\AA$), as pPXF requires the stellar templates to have a smaller FWHM than that of the data in order to accurately calculate velocity dispersion. For the instrumental FWHM, we used 2.46 $\AA$, the average of the three FWHM values found for MUSE in \citet{Raimundo2019}.

Before fitting the spectrum we masked sky lines from  N I, ($\lambda$ 5199 $\AA$), NaI ($\lambda$ 5890 and 5986 $\AA$), and OI ($\lambda$ 5577, 6300, 6364, and 6860 $\AA$). We also masked the spectrum from 7200-8450 $\AA$, as there were significant residuals from sky subtraction. 

\begin{figure*}
    \includegraphics[width=\linewidth]{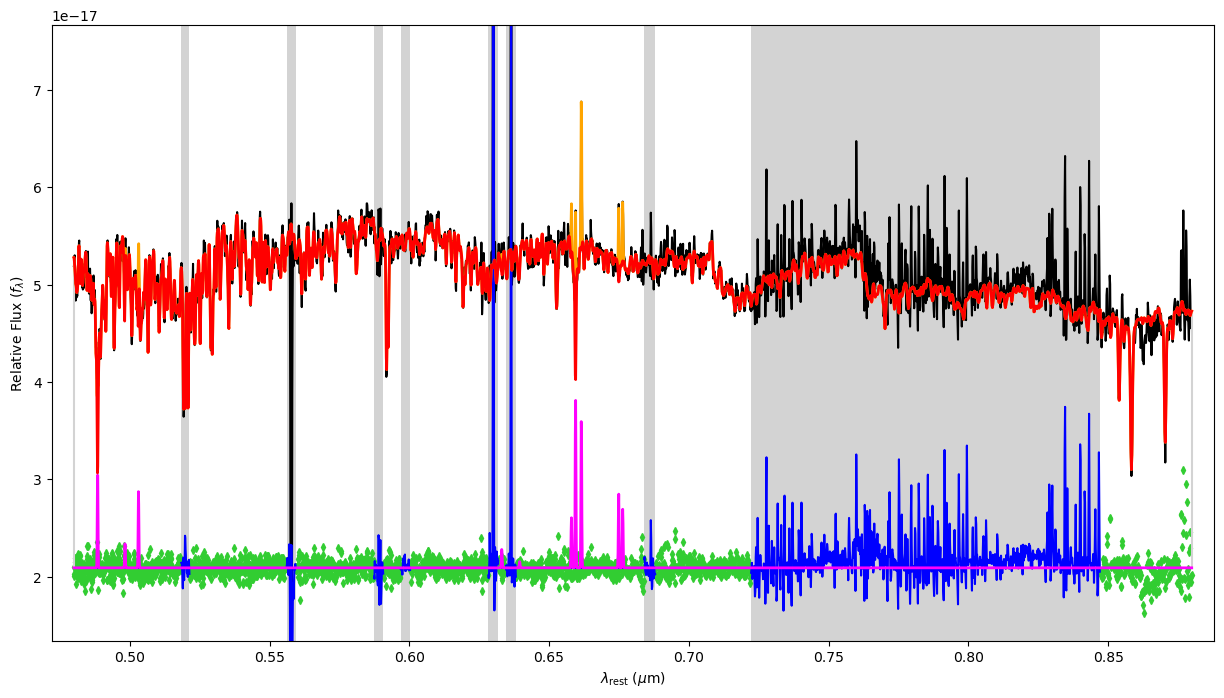}
    \caption{The integrated spectrum of the center 10 x 10 pixels, or 220 x 220 pc$^2$, of the image, fully modeled using pPXF and the methods described in Section \ref{PPXFsection}. In the top line, the black line represents the data spectrum, the red represents pPXF's model of the stellar spectrum, and orange lines model the emission lines. In the bottom portion of the image, green dots illustrate the residuals, and pink lines show the emission lines when seen without the stellar spectrum. Grey-ed out regions, and the blue lines within, are portions of the spectrum which were masked, and not considered when fitting the model. This spectrum demonstrates the power of pPXF's ability to differentiate gas emission lines from the stellar continuum. For example, in this spectrum, the stellar portion of the spectrum is found to have a deep absorption line at H$\alpha$. This enables us to measure a strong H$\alpha$ emission line, which would not have been visible in this spectrum using the naked eye. }
    \label{fig:spectrum}
\end{figure*}

To analyze each spectrum, we ran pPXF in two parts. First, it ran with a mask over all the emission lines, so that it would only be analyzing the stellar continuum. In this step, it found a weighted average of stellar templates, line of sight velocity, and velocity dispersion that best fit the data stellar spectrum. This step gave a model for the stellar continuum. Next, the mask over the emission lines was removed. Using the established stellar velocity and dispersion, as well as the weighted templates found in step one, pPXF fit Gaussians to each of the major emission lines: hydrogen transitions Balmer $\alpha$ and $\beta$, (H$\alpha$ $\lambda$6565 $\AA$ and H$\beta$ $\lambda$4862 $\AA$), [NII] (doublet around $\lambda$6583 $\AA$), [OIII] ($\lambda$5007 $\AA$), and [SII] ($\lambda$6716 and $\lambda$6731 $\AA$). With this model, PPXF calculated the line-of-sight velocity and velocity dispersion of the gas, as well as the flux of each emission line, assuming the same velocity values for all emissions lines. See figure \ref{fig:spectrum} for an example of a fully modeled spectrum. We ran this two-step process on each bin in the image, in order to create spatially-resolved maps of the emissions and dynamics across the galaxy. 
As discussed below, we then used these emission line fluxes to calculate the gas metallicity and identify sources of gas excitation.

In order to determine properties of the stellar population, we also ran a version of pPXF using the E-MILES stellar library included with PPXF \citep{Vazdekis2016}. These templates can be found at http://miles.iac.es. They have a FWHM of 2.51 $\AA$, slightly larger than the average FWHM of our data.
Thus they are not ideal for finding velocity and velocity dispersion, and we did not measure stellar velocity or dispersion in this part of the code. However, the MILES templates were sufficient to estimate the stellar population. For this step of the analysis, we masked the same sky lines as before, as well as the emission lines previously listed. The MILES library labels each stellar template with its metallicity Z = [Fe/H], a logarithmic scale where Z=0 is solar metallicity \citep{Vazdekis2010}. We edited the MILES library to only include stars with metallicity Z $>$-0.5, in order to match the range of stellar metallicities found in similar local spiral galaxies \citep{Neumann21}. pPXF then found which combination of templates best matched the spectrum for each bin, and from that combination we were able to calculate the average luminosity-weighted stellar age and metallicity. 

When processing and plotting the results from pPXF, we implemented a quality cut by masking pixels with insufficient SNR through two different methods. First, we removed any pixel where the total integrated flux was less than 0.5\% that of the pixel with the maximum total flux in the image. We then also compared the amplitude of the emission lines to the mean amplitude of noise from the binned data cube near the emission line, and removed bins where the amplitude to noise of a line (A/N) was less than five. Lastly, we masked pixels that were dominated by a star in the foreground. We show our results in Section \ref{SectionResults}.

\subsection{Modeling Disk Rotation} 
\label{ModelingDiskRotation}
pPXF allowed us to calculate line of sight velocities. We then made further investigation to determine if there were inflows of gas, a kinematically distinct core, or other deviations from simple disk rotation. 

To model the rotation of the galaxy, and deviations from a basic disk model, we used the software DiskFit \citep{SellwoodSpekkens2015}. DiskFit is a software which fits disk models to line of sight velocity maps, and allows the user to fit models varying from a basic disk to non-axisymmetric models. It does not fit spiral arms, and assumes that the galaxy has a rotationally flat central disk. DiskFit it built to fit galaxies with non-axisymmetries such as bars, using photometric or kinematic data. We used the kinematic arm of this code. DiskFit is first described in \citet{SpekkensSellwood2007}, and the most recent version of the code is explained in \citet{SellwoodSpekkens2015}.

We used DiskFit to fit models to the line of sight velocity maps that we found for the gas and stars in this galaxy. For the input data, we used the velocity maps for gas and stars that were outputted by pPXF. Pixels with total integrated flux less than 0.5\% of the maximum flux in the image were set as "not a number," so that they would not be considered in the fit. We fit several models to these data, including a simple disk, radial flows, warping of the disk, and non-axisymmetric flows, in order to find the models that produced the lowest minimum $\chi^2$ value. We show the results in Section \ref{sec:DiskFitResults}.

For each model we allowed DiskFit to fit the position angle, ellipticity, disk center, inclination, and systematic velocity for each model. We set the seeing correction at 3 pixels, corresponding to the FWHM of our data. We also held several parameters constant in each model, including the starting guesses for parameters, max delta velocity, and the region fit. Another parameter held constant for each model was delta ISM, a parameter in DiskFit's kinematic code that accounts for the velocity uncertainty associated with the turbulence in the interstellar medium. We set this value to 10 km/s for gas and 5 km/s for stars, as gas dynamics are more strongly affected by turbulence. We allowed the software to assume that the velocity would rise linearly from 0 at the center of the galaxy, because spaxels close to the center have a high signal level. DiskFit used all of the pixels within a central ellipse with a semi-major axis of 100 pixels, with an ellipticity of 0.3. Outside of this ellipse, DiskFit evaluated pixels in rings, stepping out by six pixels for each ring (in order to be double the seeing correction), out to a radius of 165 pixels. When fitting non-circular flow features, we used DiskFit to examine pixels from radius = 3 to 105, as 105 pixels ($\sim$2.3 kpc) was both where we estimated the galactic bar to reach, and the largest radius that caused the non-axisymmetric model to give physically plausible results. 

Once we had created these basic models, we evaluated the residuals and minimum $\chi^2$ value for each model, in order to determine which model best fit our data. We found that the model with m=2 non-axisymmetric flows, which we will refer to as the bar model, was the best fit for both stars and gas. (See section \ref{SectionResults}). From here, we fit the basic disk and bar models to the gas and stars velocity data, this time using bootstrapping, a statistical method in which we fit the model many times, slightly varying the parameters each time, in order to determine uncertainties. We used the bootstrap technique described in \citet{SZS10}, with 100 bootstrap iterations. This allowed us to calculate the uncertainties of the parameters output by DiskFit. 

\section{Results}
\label{SectionResults}
\subsection{Spatial Distribution of Stars and Gas}
With the results from pPXF, we created two-dimensional maps of the dynamics and gas emissions of the galaxy. 

\begin{figure*}[ht]
    \includegraphics[width=.5\linewidth]{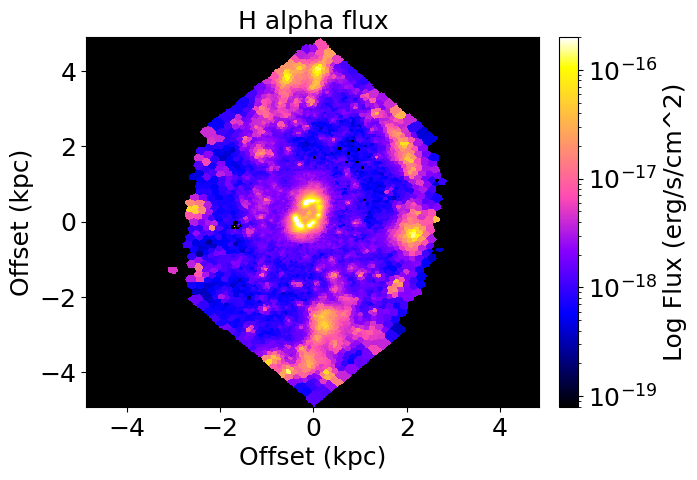}\includegraphics[width=.5\linewidth]{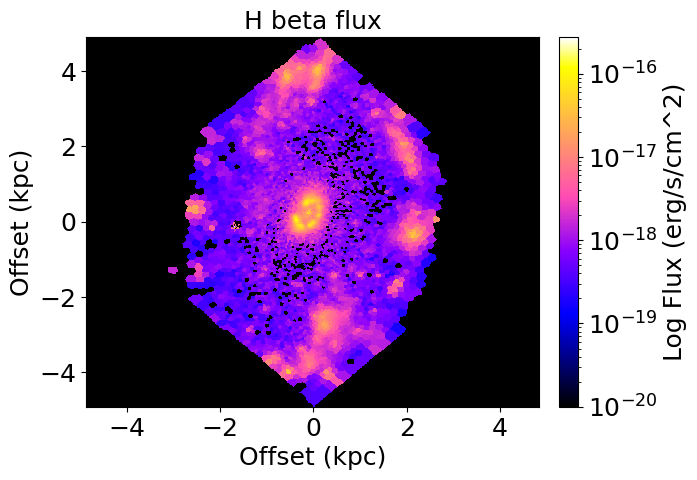}
    \includegraphics[width=.5\linewidth]{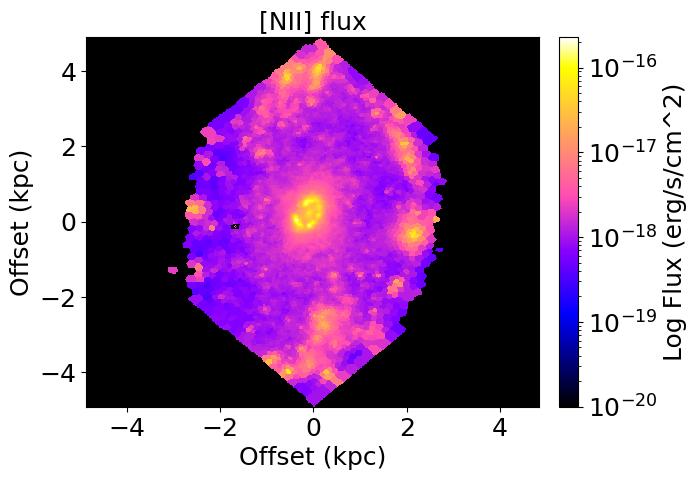}\includegraphics[width=.5\linewidth]{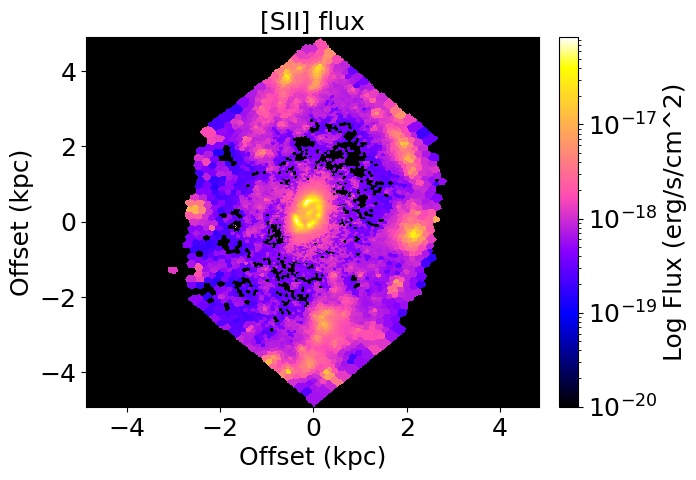}
    \includegraphics[width=.5\linewidth]{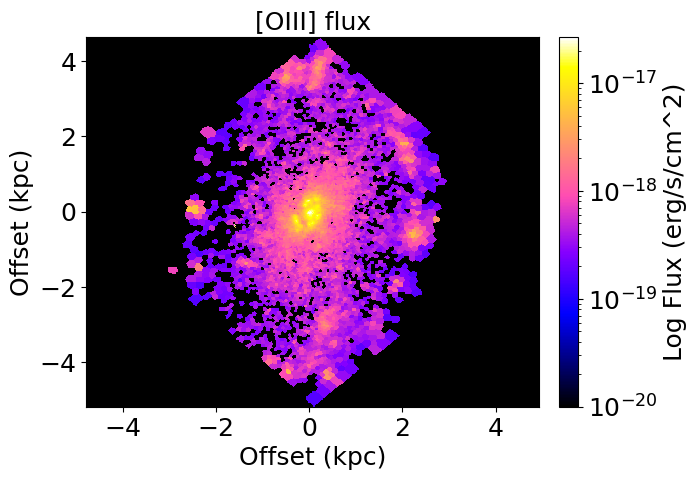}
\label{emissionsfig}
\caption{Two dimensional flux maps of the major emission lines observed in the optical range: H$\alpha$, H$\beta$, [NII], [SII], and [OIII]. The black pixels in these images are bins which were removed because the signal to noise ratio of the emission line was lower than 5, or the total flux was less than 0.5\% of the maximum flux of the image. In these and all other images created from the MUSE data cube, North is up and East is left.}
\end{figure*}

\begin{figure*}[ht]
    \includegraphics[width=.5\linewidth]{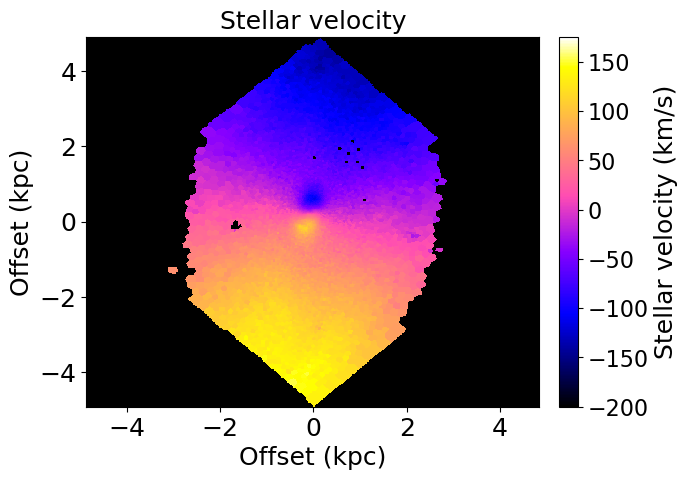}\includegraphics[width=.5\linewidth]{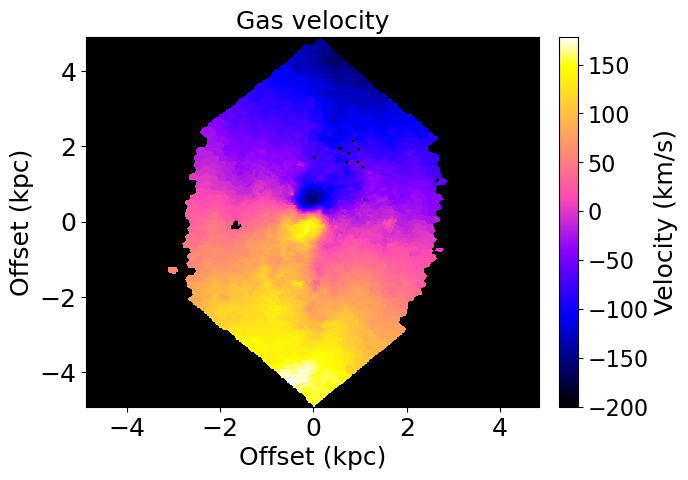}
    \includegraphics[width=.5\linewidth]{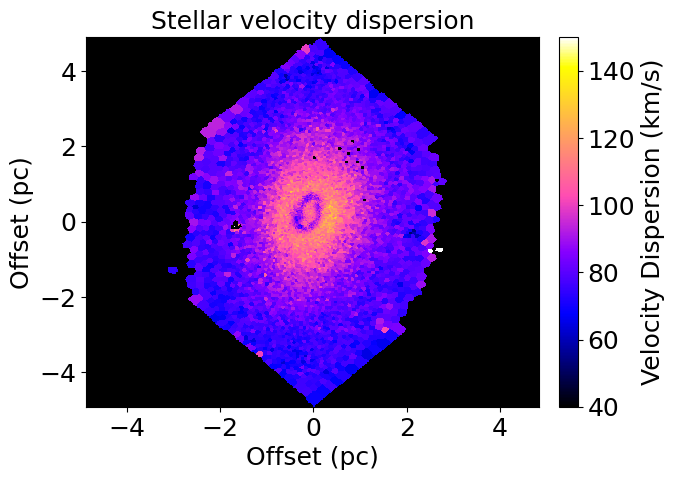}\includegraphics[width=.5\linewidth]{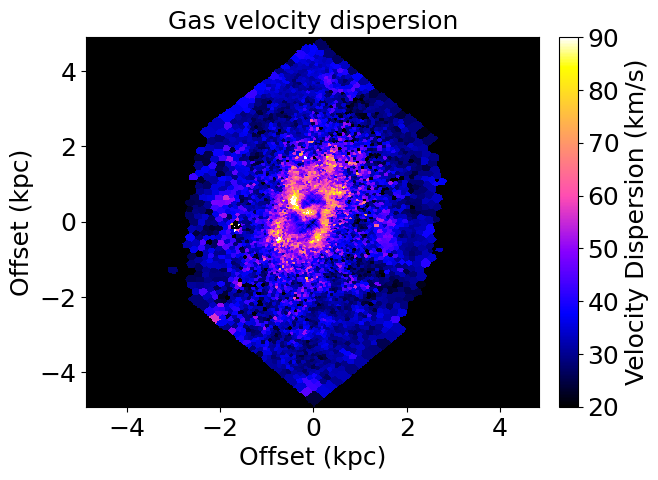}
\caption{Two dimensional maps of stellar (left column) and gas (right column) dynamics. Line of sight velocity is shown at the top and velocity dispersion at the bottom. North is up, East is left. The small masked region in black, on the left-hand side of each map, corresponds to a star in the foreground, which was removed from all final maps.}
\label{BasicDynamics_fig}
\end{figure*}

Figure \ref{emissionsfig} includes two-dimensional  flux distributions for several major emission lines that trace the ionised gas of the galaxy: H$\alpha$, H$\beta$ [NII], [OIII], and [SII]. In all of theses images, there is a bright ring around the center of the galaxy, with a projected radius of $\sim$400 pc. Emission maps also show a larger spiral arm with clumpy bright regions. We observe bright emissions in the very center of the galaxy, especially in [OIII]. 

Figure \ref{BasicDynamics_fig} displays the maps of dynamics in the galaxy. Due to our 2-part method of running PPXF, we were able to separate the dynamics of the gas from the stars. Figure \ref{BasicDynamics_fig} shows the stellar line of sight velocity, the stellar velocity dispersion, ionised gas line of sight velocity, and gas velocity dispersion.

The gas velocity maps show departures from axisymmetry, which can be seen in the zero-velocity line illustrated in figure \ref{fig:gasvel_sshape}.
Additionally, the circumnuclear ring which was seen in the emission maps, in the central 800 pc of the image, also appears in the dynamics maps. Here we see that both the stellar and gas components of the ring are rotating around the same axis as the main stellar body, but more rapidly than the ring's surroundings. 
The dynamics plots also show that the stellar velocity dispersion of the inner ring is lower than that of its surroundings, suggesting the presence of a dynamically cold disk of relatively young stars. The gas velocity dispersion is similarly low within the ring. Just outside the central ring, the gas velocity is significantly greater than it is in the rest of the image. This enhanced gas velocity dispersion reaches farther away from the ring in the north-west direction and south-east. The increased gas velocity dispersion may indicate that the AGN is inducing turbulence in the gas just outside the circumnuclear ring. 

\subsection{BPT Diagrams}
\begin{figure*}
    \includegraphics[width=.5\linewidth]{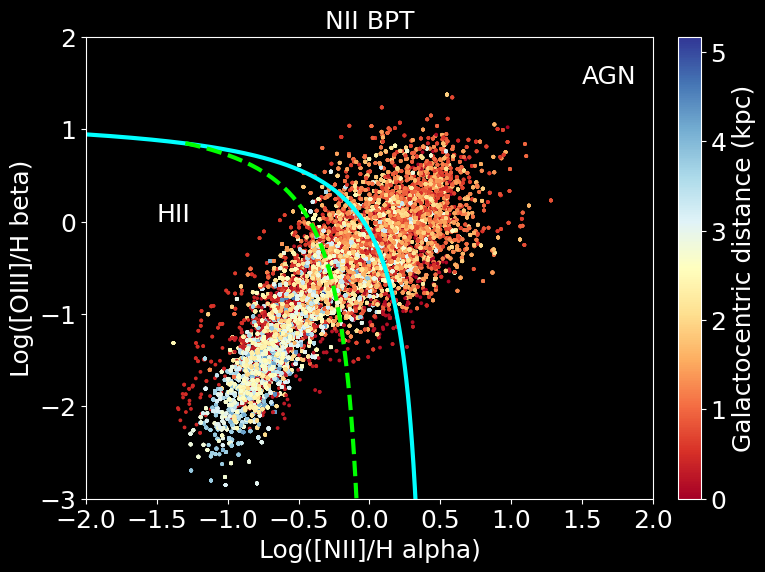}\includegraphics[width=.5\linewidth]{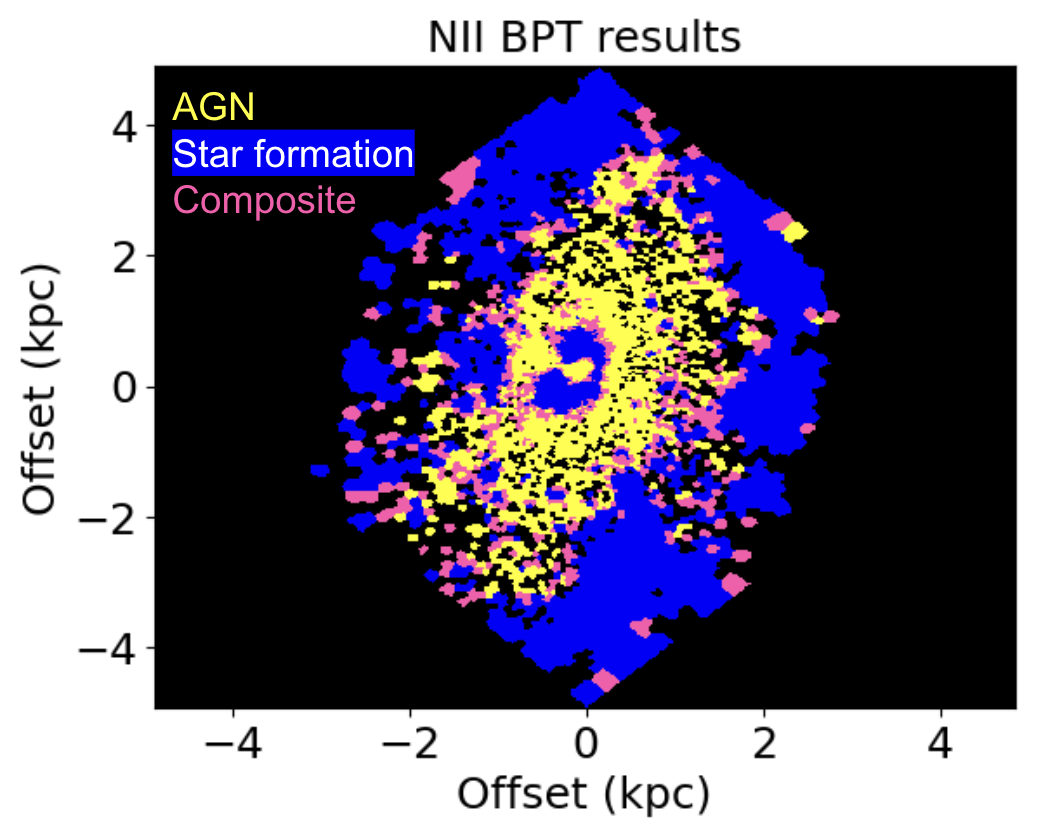}
\caption{(Left) The NII BPT diagram, with each spaxel plotted individually, using emission line ratios to categorize each bin in the image by its main source of gas excitation. Points that fall to the right of the solid blue line indicate gas dominated by AGN excitation, while points to the left of the dashed green line indicate an HII region. Points that fall between the two lines are classified as composite. The colorbar indicates the spaxel's distance from the center of the galaxy, in kpc. (Right) The results from the BPT diagram--that is, which section of the BPT diagram each spaxel lands within--are then mapped on the right panel over the image of the galaxy. Yellow pixels indicate AGN excitation, blue pixels mark star-forming HII regions, and pink pixels are composite. }
\label{BPTfig}
\end{figure*}

We use the commonly-used [NII] Baldwin,
Phillips and Terlevich (BTP--\citet{Baldwin1981}) diagram to identify regions where gas excitation is primarily dominated by AGN excitation, radiation from newly formed stars in HII regions, or a combination thereof \citep{Baldwin1981,Kewley2006}. In this diagram, we plot the [OIII]/H$\beta$ flux ratio as a function of the [NII]/H$\alpha$ flux ratio, on the plot seen in Figure \ref{BPTfig} (left). The solid blue line represents the extreme starburst boundary line from \citet{Kewley2001}, while the dashed green line shows the \citet{Kauffmann2003} pure star-formation track. Points that fall above and to the right of both of these lines are classified as AGN-dominated, while points to the the left and below the \citet{Kauffmann2003} line are marked as star-formation dominated. Points in between the two lines are classified as composite, potentially with an excitation contribution from the AGN, star formation, shocks, or a combination thereof. 

Using the emission lines previously discussed, we plotted each bin of the image on a [NII] BPT diagram, and then mapped the spatially resolved results back to the physical regions of the galaxy (see Figure \ref{BPTfig}, right). Pixels were removed from this map if any of the four emission lines used did not each have the required SNR and total flux threshold, as discussed in section \ref{PPXFsection}. This BPT map indicates that the the inner circumnuclear ring and the outer spiral arm are HII regions, primarily ionized by star formation. It also shows that gas at the center of the galaxy, and in the intermediate region, is primarily excited by the AGN. This excitation from the AGN reaches up to 3500 pc away from the center of the galaxy. This is the same trend seen in other spatially resolved maps of line emission in Seyfert 2 galaxies \citep{Xia2018}. In the case of high OIII/H$\beta$  (Seyfert) regions, the mechanism is believed to be photoionization by the hard AGN continuum photons. In the case of the high [NII]/Halpha (LINER) regions, it is possible the energy transfer is mechanical, perhaps in the form of an outgoing wind. 

\subsection{Gas metallicity}
\begin{figure*}
    \includegraphics[width=.5\linewidth]{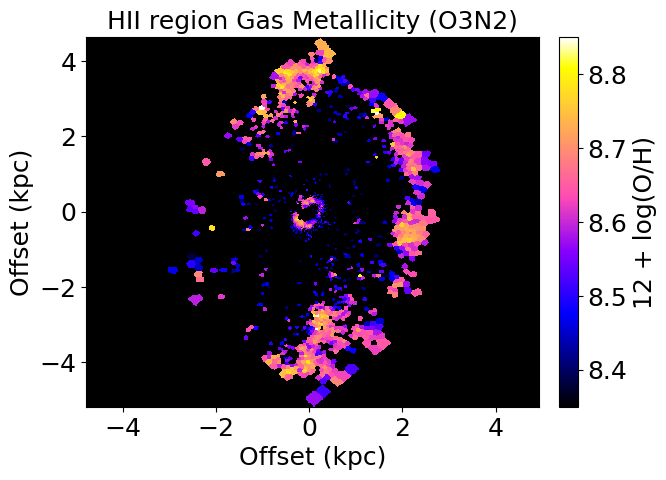}\includegraphics[width=.5\linewidth]{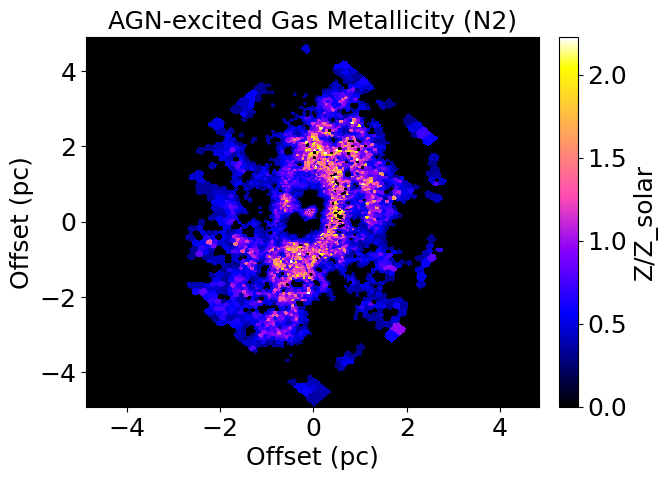}
\caption{(Left) Gas metallicity calculated using the O3N2 callibration for HII regions from \citet{Marino2013}, mapped in BPT-identified HII regions. The metallicity remains close to solar values throughout the image, with no visible gradient. (Right) Metallicity of gas in regions excited by the AGN, using the N2 calibration for AGN regions from \citet{Carvalho2020}. }
\label{Gasmetalfig}
\end{figure*}

Next we calculated the gas metallicity, from the [OIII]/H$\beta$ and [NII]/H$\alpha$ line ratios and the O3N2 metallicity calibration for HII regions from \citet{Marino2013}.  We calculated the O3N2 metallicity of the inner and outer rings, where the BPT diagram indicates that there are HII regions (Figure \ref{Gasmetalfig}, left). In these images, as in the BPT maps, we remove pixels where any of the diagnostic emission lines have SNR $<$ 5. We also remove any pixels where the Balmer decrement is measured to be lower than 2.86, the theoretical Balmer ratio \citep{OsterbrockFerland2006}, to avoid pixels where the Balmer decrement is unphysical. We also calculated the metallicity of gas in regions excited by the AGN, as indicated by the BPT map, using the [NII] calibration from \citet{Carvalho2020}, which was calibrated for AGN-excited regions (Figure \ref{Gasmetalfig}, right). 

The results from both calibrations indicate that there is no significant metallicity gradient in the disk of this galaxy. In the HII regions of the central ring and the outer spiral, the gas metallicity stays at and slightly above the solar metallicity, 8.69 \citep{Asplund2009}. Similarly, in the central region, the gas metallicity is near solar metallicity. The gas metallicity is slightly higher in the intermediate region, but never goes above 2 times solar metallicity. Figure \ref{Gasmetalfig} shows maps of gas metallicity throughout the image, with the HII calibration on the left and the AGN calibration on the right. 

\subsection{Average stellar population}
\begin{figure*}
    \includegraphics[width=.5\linewidth]{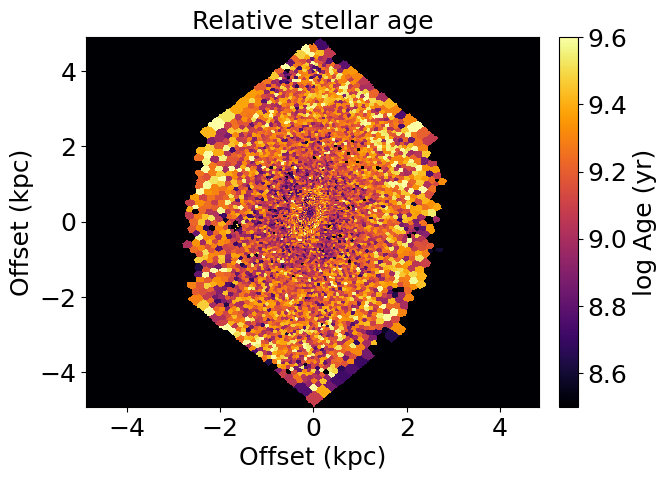}\includegraphics[width=.5\linewidth]{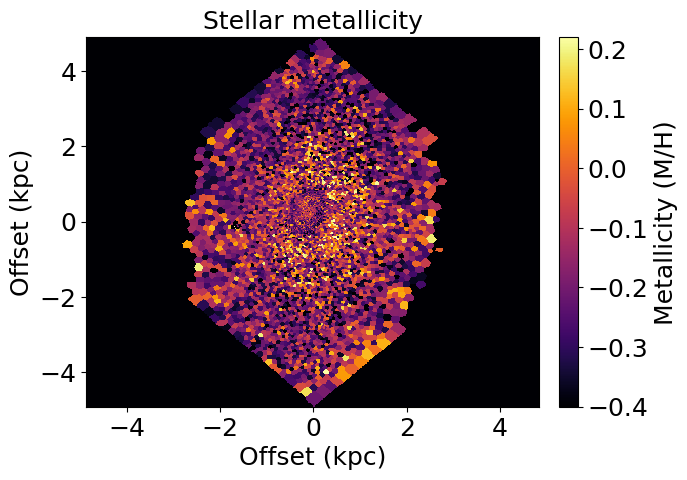}
\caption{(Left) Average stellar age. (Right) Average stellar metallicity. As in the full gas and stellar fit, we removed pixels where the total integrated flux was less than 0.5\% of the max integrated flux in the image. The stars in the central ring and spiral arm are somewhat relatively old and low-metallicity, while the youngest and highest-metallicity stars seem to appear in the central and intermediate regions.}
\label{starpopfig}
\end{figure*}

To determine the average properties of the stellar population, we use the results from pPXF to map the average stellar age and metallicity, displayed in Figure \ref{starpopfig}. These images show that the central ring is primarily made up of relatively old, low-metallicity stars. This ring is surrounded by a region of younger, higher-metallicity stars. This ring of relatively old stars is unexpected, as it is co-spatial with regions that the BPT diagram marked as locations of star formation. 

In order to test the statistical significance of the stellar population results, we ran the central ring and the intermediate region through a Monte Carlo simulation. In this simulation, we ran PPXF on the spectrum of each pixel 200 times, varying the starting parameters slightly each time, in order to find the uncertainties of the output parameters. The results of this Monte Carlo test are listed in table \ref{table1}. We find that there is a $> 1 \sigma$ difference in age and metallicity between the circumnuclear ring and the surrounding area. Therefore, we tentatively find that the stars in the central ring are relatively old and low-metallicity, but these results are not statistically significant. 

\begin{table}[]
\begin{tabular}{|l|l|l|}
\hline
                 & Central ring     & Intermediate region        \\ \hline
Mean age (log yr)   & 9.43 $\pm$ 0.02  & 9.08 $\pm$ 0.20  \\ \hline
Mean metallicity (M/H) & -0.33 $\pm$ 0.01 & -0.10 $\pm$ 0.16 \\ \hline
\end{tabular}
\caption{The mean stellar age and metallicity in the central ring and the intermediate region. Both regions were run through a Monte Carlo simulation to test the statistical significance of the findings. The stellar population of the central ring seems to be older and lower-metallicity than the population of the intermediate region, but these results are within $1 \sigma$ of each other and thus not statistically significant.}
\label{table1}
\end{table}

\subsection{Modeling Dynamics with DiskFit}
\label{sec:DiskFitResults}

\begin{figure*}[]
    \includegraphics[width=\textwidth]{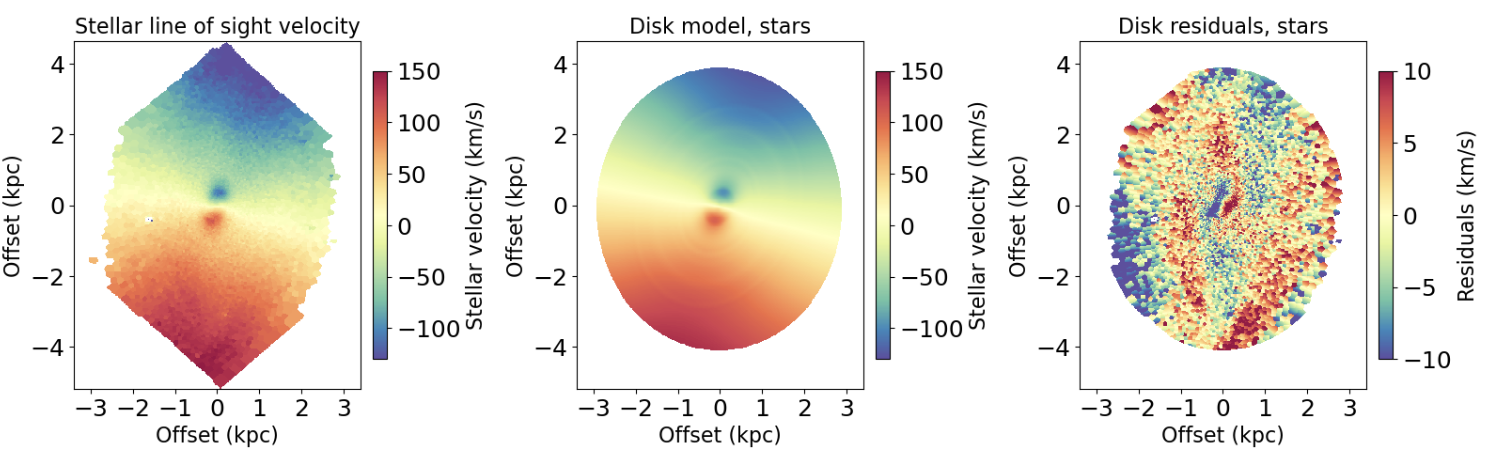}
    \includegraphics[width=\textwidth]{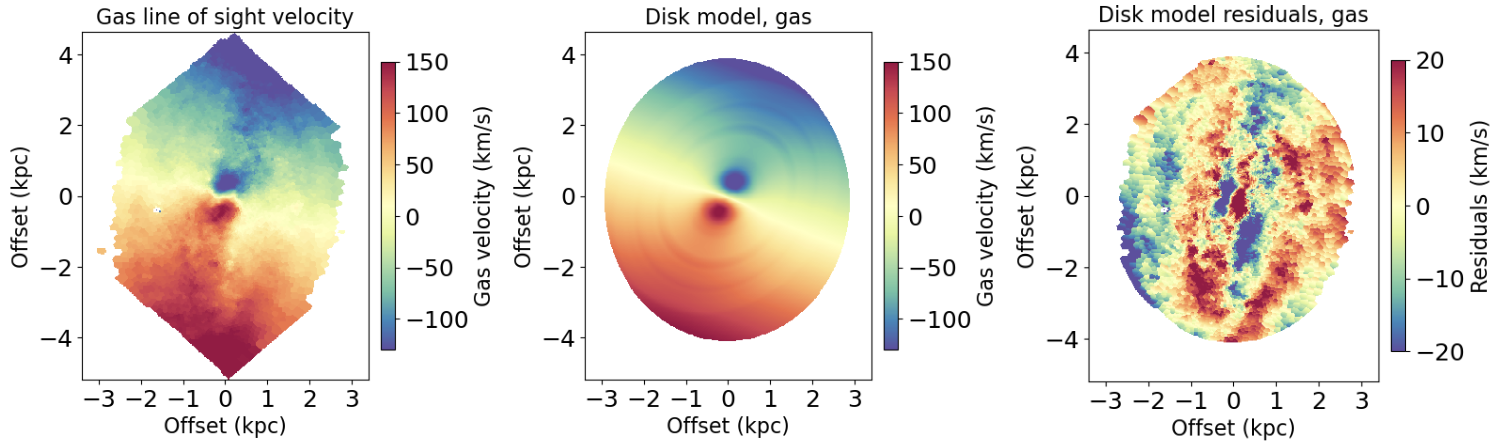}
\caption{Maps of DiskFit modelling, fitting a basic disk model to the stellar and gas velocity. The top row, from left to right, includes the stellar velocity map output by PPXF, a model fitting this velocity data as a basic disk, and the residuals from subtracting this model from the data. The bottom row includes the same maps, but for gas rather than stars.}
\label{DiskFit:BasicDisk}
\end{figure*}

\begin{figure}
    \includegraphics[width=0.5\textwidth]{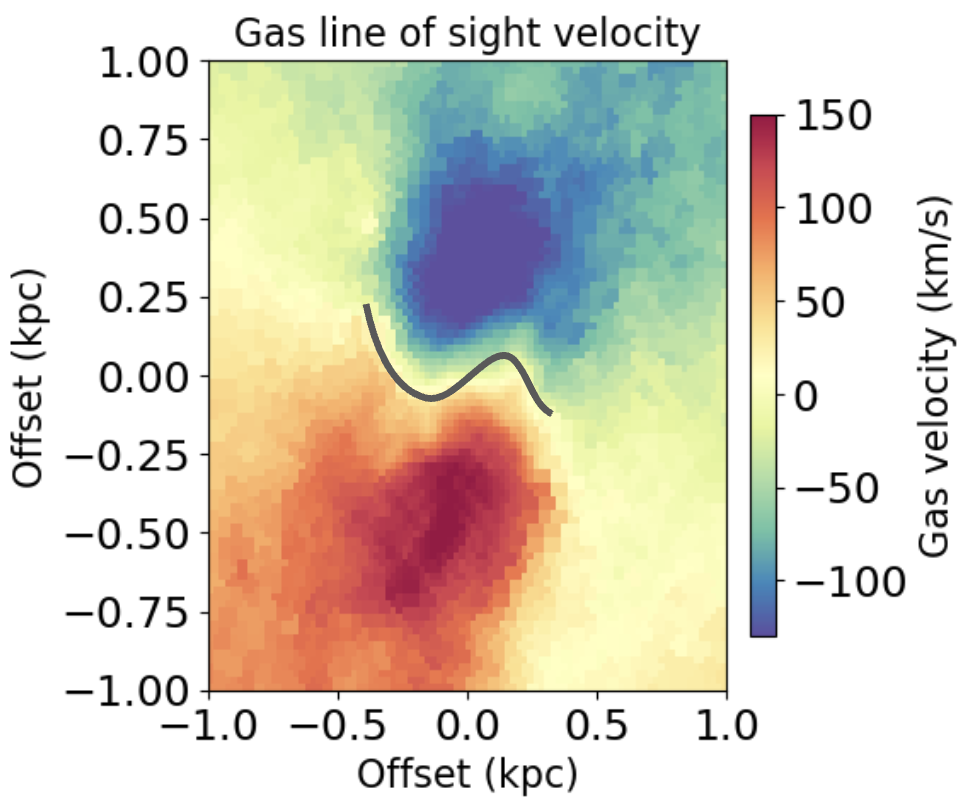}
\caption{Line of sight velocity of gas in the center 1 x 1 kpc$^2$ of the galaxy, with the s-shaped distortion illustrated. This distortion indicates the presence of a galactic bar.}
\label{fig:gasvel_sshape}
\end{figure}

In order to better understand the dynamics of this galaxy, we used the software DiskFit to fit several models to our stellar and gas velocity maps. (See Section \ref{ModelingDiskRotation} for more details on DiskFit.) The first model that we tried was basic disk rotation. In Figure \ref{DiskFit:BasicDisk}, we display the stellar and gas line of sight velocities, basic disk model, and residuals. In the residuals from this basic model, for both the stars and gas, we see large, anti-symmetrical residuals in the central 1 x 1 kpc$^2$ of the galaxy. (See the right-most panels in both rows of figure \ref{DiskFit:BasicDisk}). This type of residuals has been found by \citet{LopezCoba22} to indicate the presence of a galactic bar. We also see an S-shaped distortion in the line-of-sight velocity map for gas, illustrated in figure \ref{fig:gasvel_sshape}. This s-shape was also identified by \citet{LopezCoba22} as a signature of bar-driven kinematics.  Therefore, we explored several other models available through DiskFit, in order to find a better model for our data.

\begin{table}[]
\begin{tabular}{|l|l|l|}
\hline
            & Stars             & Gas               \\ \hline
Disk PA     & -12.2 $\pm$ 0.5 & -19 $\pm$ 1 \\ \hline
Ellipticity & 0.41 $\pm$ 0.02   & 0.28 $\pm$ 0.04   \\ \hline
Inclination & 54 $\pm$ 1  & 44 $\pm$ 3  \\ \hline
Bar PA      & 104 $\pm$ 4  & 69 $\pm$ 5  \\ \hline
\end{tabular}
\caption{Best-fit parameters from DiskFit for a model with a galactic bar, modeling velocity maps of gas and stars. Position angle and inclination are in units of degrees. The ellipticity and the position angle (PA) of the disk and the bar are notably different for the gas and stars. }
\label{tab:table2}
\end{table}

\begin{figure*}[]
    \includegraphics[width=\textwidth]{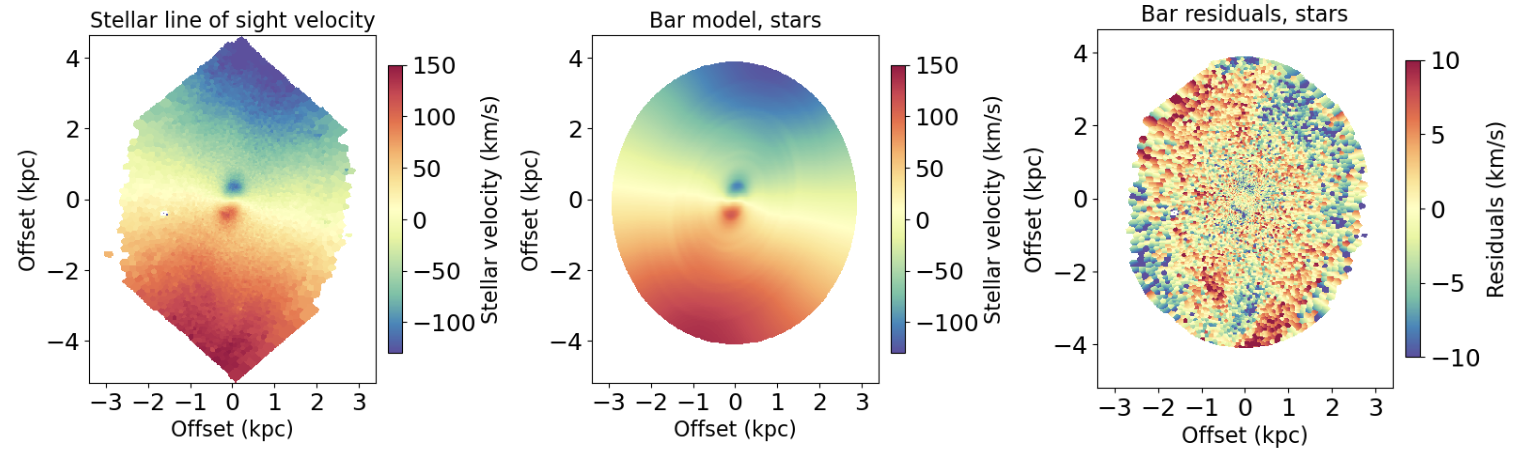}
    \includegraphics[width=\textwidth]{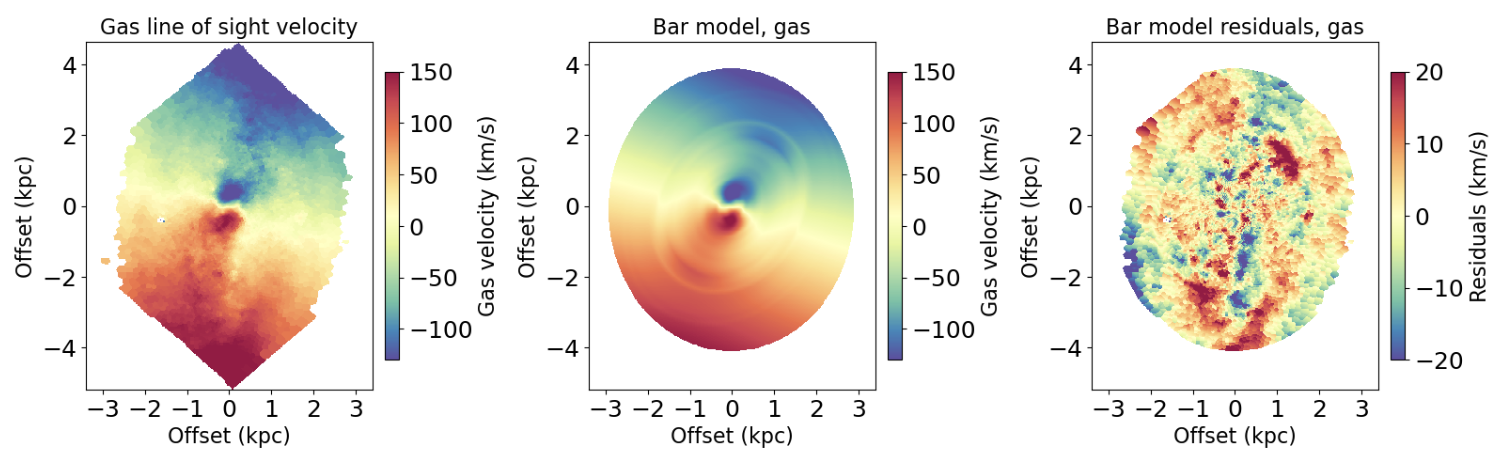}
\caption{Maps of DiskFit modelling with the bar model. This model fits the data better than the basic disk model. The top row shows the stellar line of sight velocity, bar model, and the residuals from this model. The bottom row contains the same maps, but for gas.}
\label{DiskFit:NAS}
\end{figure*}

After the basic disk showed signs of non-axisymmetry, we tried several different models: a warped disk, radial flows, and non-axisymmetric flows with m=1 and m=2. The non-axisymmetric flow patterns are modeled using summations of cos(m$\theta$); for more detail see \citet{SpekkensSellwood2007}. We found that a disk with non-axisymmetric flows at harmonic order m=2 was the model that best fit the data and produced the lowest minimum $\chi^2$ value, for both gas and stars. This model was built to fit barred spiral galaxies, and thus we will refer to it as the bar model \citep{SellwoodSpekkens2015}. (See figure \ref{DiskFit:NAS} for illustration of the bar model).
Therefore, in agreement with the residuals that we observed in the basic disk model, the non-axisymmetric distortion in NGC 5806 is best explained by the presence of a bar. With this model, we found the parameters listed in table \ref{tab:table2}.

The bar model fits the observations significantly better than the basic disk model. This can be seen in the reduced structure in the residuals of the bar model, as compared to the anti-symmetric residuals from the basic disk model (see Figures \ref{DiskFit:BasicDisk} and \ref{DiskFit:NAS}). The bar model also has lower $\chi^2$ values than the basic disk model, for both gas and stars, as shown in table \ref{chitable}. We limited the length of the bar to 105 pixels, as this was what we estimated using the Hubble image in Figure \ref{HubbleCompositeImage},
and this size of bar--2.3 kpc in radius--gave the best fit for the DiskFit model. There is still some structure in the residuals of the stars and gas velocity, with positive residuals in the top-left and bottom-right side of the image, and negative residuals in the top-right and bottom left; however, this can be explained by the presence of the spiral arm, which the DiskFit model does not include.

\begin{table}[]
\begin{tabular}{|l|l|l|}
\hline
             & Stars & Gas  \\ \hline
Basic Disk   & 1.22  & 1.83 \\ \hline
Bar          & 0.88  & 0.85 \\ \hline
\end{tabular}
\caption{$\chi^2$ values for the basic disk model and bar model, when compared to line of sight velocity maps for the gas and stars of this galaxy. These $\chi^2$ values enable us to compare the bar model to the disk model, and show that the bar model is a better fit.}
\label{chitable}
\end{table}

In addition, we ran DiskFit on the inner 2 x 2 kpc$^2$ of the galaxy, in order to test if the inclination or position angle of the central ring might differ from that of the major disk of the galaxy. We found no such discrepancy; the central ring is aligned with the major gas disk of the galaxy.

\subsection{Maps of the central ring}

\begin{figure*}
    \includegraphics[width=.33\linewidth]{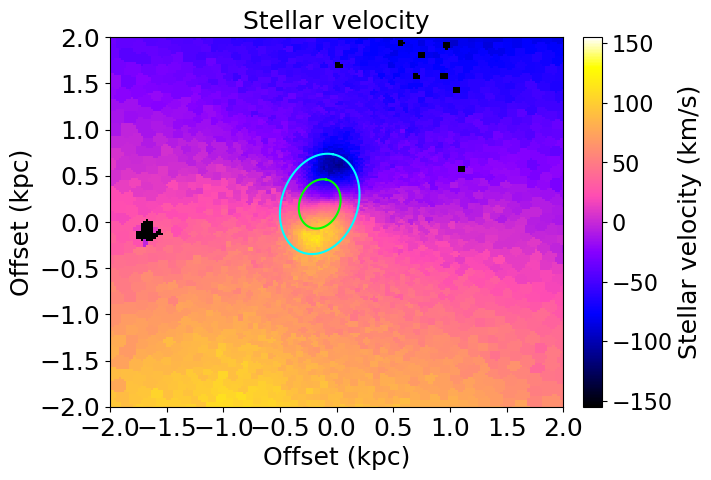}\includegraphics[width=.33\linewidth]{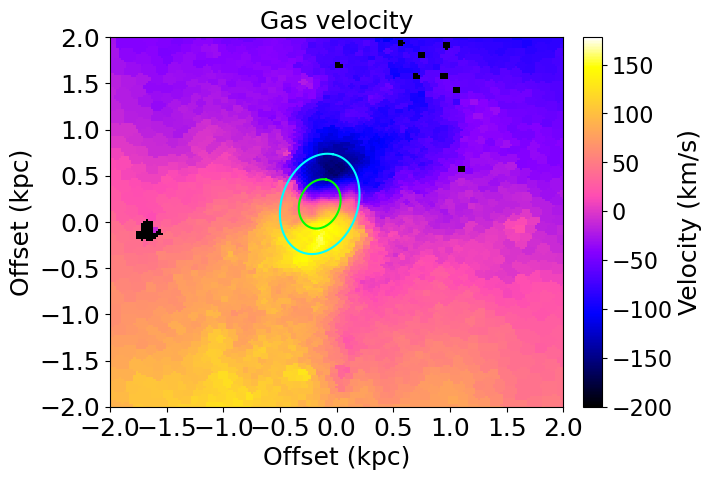}\includegraphics[width=.33\linewidth]{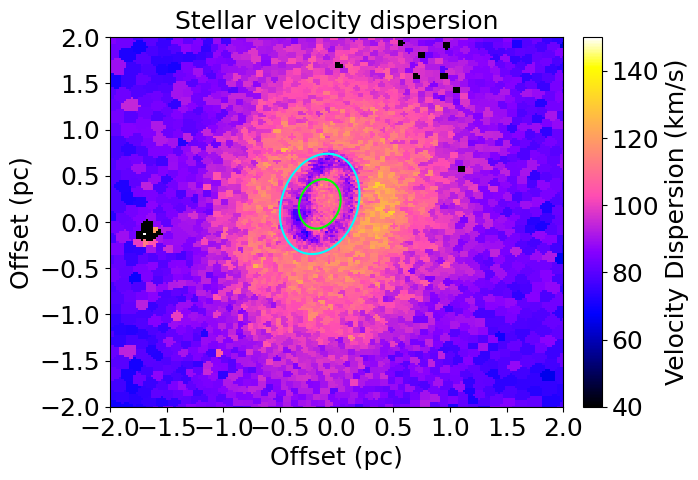}
    \includegraphics[width=.33\linewidth]{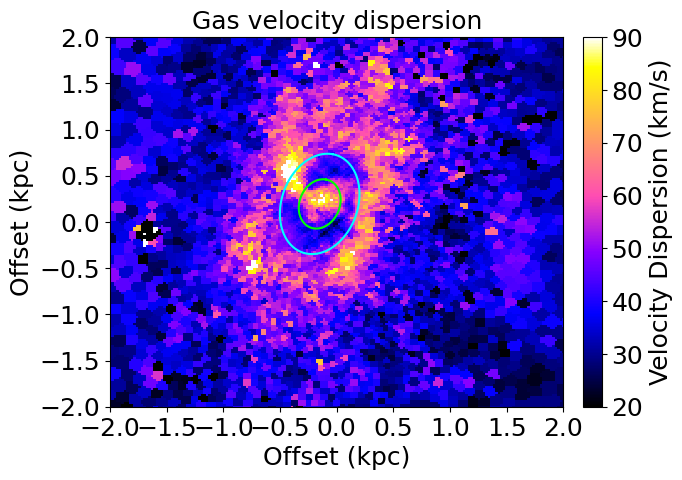}\includegraphics[width=.33\linewidth]{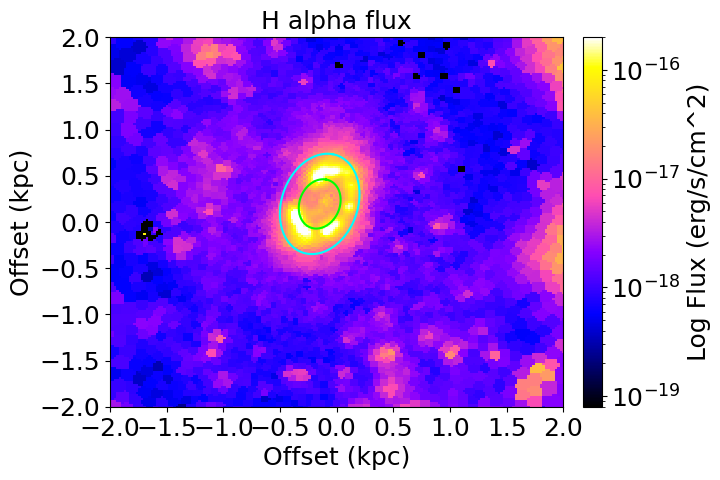}\includegraphics[width=.33\linewidth]{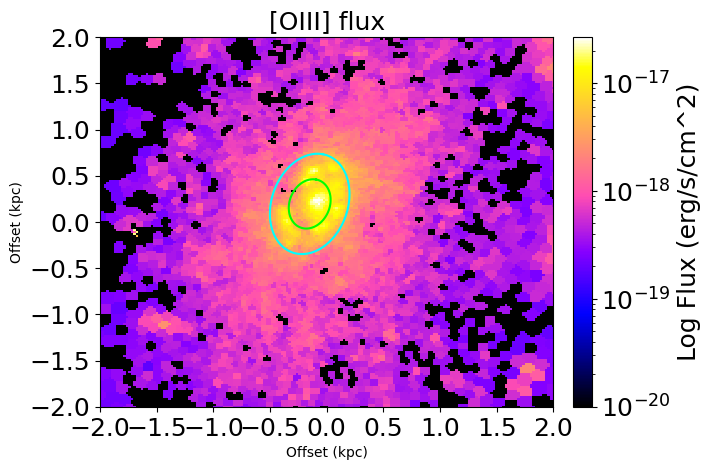}
    \includegraphics[width=.33\linewidth]{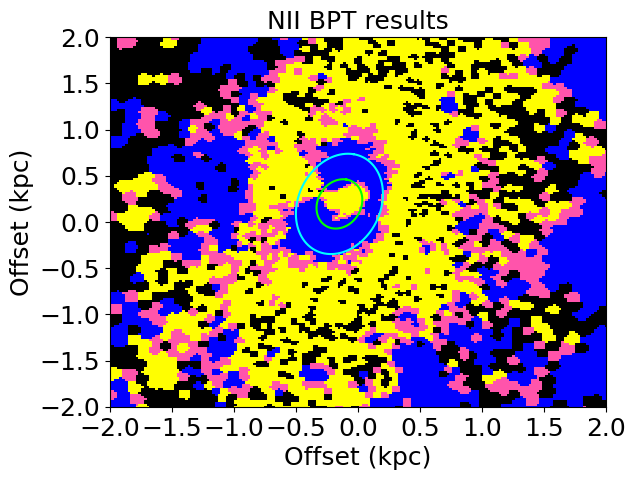}\includegraphics[width=.33\linewidth]{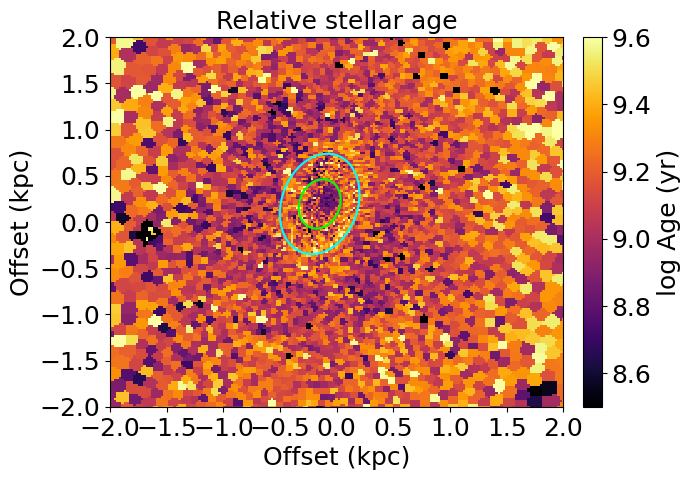}\includegraphics[width=.33\linewidth]{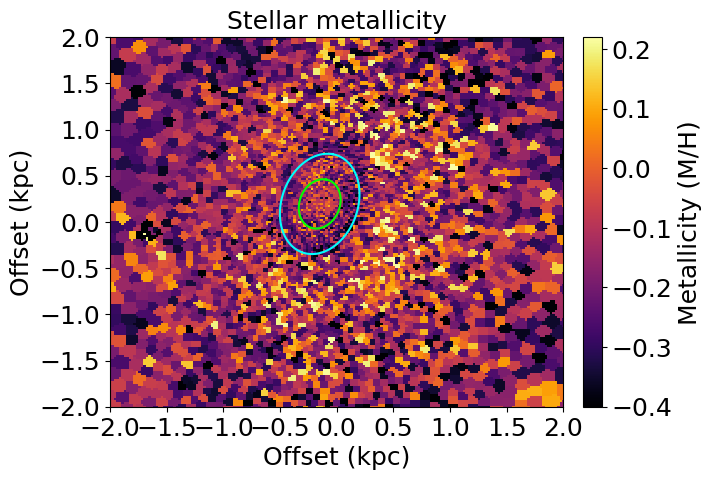}
\caption{Maps of stellar velocity, gas velocity, stellar velocity dispersion, gas velocity dispersion, H$\alpha$ flux, [OIII] flux, results from the BPT diagram, average stellar age, and average stellar metallicity, all zoomed in on the central 2 kpc$^2$ of the galaxy. Ellipses are superimposed in the same location in each map, in order to demonstrate that all of the effects that we observe are co-spatial.}
\label{CenterRingFig}
\end{figure*}

We present the maps previously featured, zoomed in on the central ring, in Figure \ref{CenterRingFig}. The blue ellipses are superimposed so that the viewer can compare the location of the ring in each image. These images show that the ellipse visible in each map is in fact the same central ring. This indicates that the maximum gas emissions and oldest and lowest metallicity stars in the image all occur in the same space, which has been identified as HII regions. This ring is also a location of high stellar and gas velocity, and low stellar and gas velocity dispersion. Normally, this low velocity dispersion would indicate the presence of newly-formed stars, as would be expected in HII regions; however, our analysis of the stellar population instead indicates that the average stellar age in this ring is relatively old compared to the regions around it. This ring has a projected radius of about 400 pc. Potential reasons for this discrepancy will be discussed in the discussion section below. 

\section{Discussion}
\label{SectionDisc}
\subsection{AGN-dominated emissions}
With the quality of data provided by MUSE, we have been able to measure the gas excitation mechanisms of different regions of this galaxy with great detail. \citet{Erroz-Ferrer2019} found that the central region of this galaxy showed signs of AGN excitation, a finding which we have confirmed. In addition, due to our method of increasing the SNR through Voronoi binning, we also find significant AGN emissions in the intermediate region, reaching over 3 kpc away from the AGN. The reach of these emissions suggests that the AGN may be affecting the gas in this galaxy up to greater distances than previously observed. 

\subsection{Central ring}
In the center of this galaxy, we see a star-forming central ring with a projected radius of approximately 400 pc. This star-forming ring has been observed by several previous studies \citep{Carollo2002,Dumas2007,Erroz-Ferrer2019,CLB2023}. We find that this ring has a high concentration of excited gas, and lies in the HII region zone on a BPT diagram.


We have also shown that this ring rotates distinctly from the rest of the galaxy, with rapid rotation and low velocity dispersion.
These dynamics show elevated rotational support consistent with a dissipative formation channel, such as gas inflow accumulation caused by a bar. In addition, the ratio of the ring to bar size falls within ranges previously found for nuclear disks fueled by bars. The size of the ring is $\sim$0.17 times the radius of the bar. This is slightly larger than the values found by \citet{Gadotti2020} for the typical sizes of nuclear disks formed by bars, and slightly lower than the predicted theoretical value of 0.3 \citep{Cole2014}. 

In all of the emission line maps that we have created, we see that there is a gap in the north-east portion of the ring. Based on the Hubble image in Figure \ref{HubbleCompositeImage}, we believe that this portion of the ring is merely obscured by a dust lane, which would explain the gap observed. In order to test this, we found the amplitude of the H$\alpha$ and H$\beta$ lines within an integrated spectrum for this region. We calculate the Balmer decrement in the gap in the ring to be 7.3. This falls on the high end of Balmer decrement values measured by \citet{Lu2019} in the central regions of active galaxies. Therefore we find that there is a significant amount of dust in this portion of the ring.

\begin{figure*}[ht]
\centering
  \includegraphics[width=.8\linewidth]{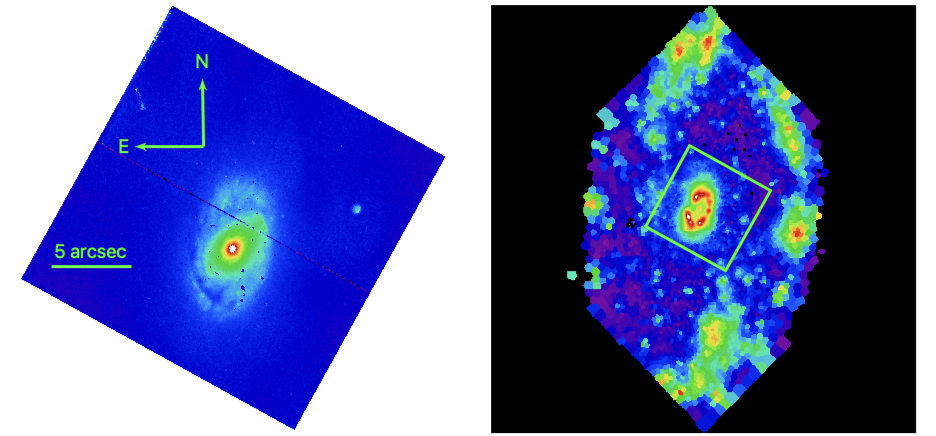}
\caption{Left: Infrared image of the center of NGC 5806, taken by Hubble Space Telescope's Near Infrared Camera and Multi-Object Spectrometer (NICMOS). This image was taken with NICMOS' F110W filter, on February 2, 1998 (HST proposal ID $\#$7331, PI: M. Stiavelli; \cite{Carollo1998}). There is a $\sim$3 arcsecond offset in coordinates between this NICMOS image and our MUSE data. The brightest point in this image is the AGN, and the ellipse pictured is co-spatial with the central region and circumnuclear ring seen in the optical. Right: H$\alpha$ map from MUSE data cube, with the field of view of the IR image superimposed in green for scale reference.}
\label{IRim}
\end{figure*}

To view this region without the dust lane blocking our view, we reference the infrared image displayed in Figure \ref{IRim}. This image was taken by Hubble Space Telescope's Near Infrared Camera and Multi-Object Spectrometer (NICMOS), in the central R$\approx$15 arcmin of NGC 5806. There is a $\sim$3 arcsecond offset in coordinates between this NICMOS image and our MUSE data, as can be observed by interpreting the brightest point in the IR image to be the AGN. This image reveals a reservoir of dust within the central region, which may contribute to feeding the AGN. Another interesting feature of this image is that the southern portion of the central ring resembles a spiral arm, which could potentially warrant further investigation.

More pertinently to this portion of the discussion, the NICMOS image shows the circumnuclear ring to be largely unbroken in the northeast. Therefore, it is likely that the physical ring does not have a gap in it. The gap in optical emission, however, could still be of interest.
The gap in optical emission could be in part due to the inflow of gas disrupting star formation. At the same location where we see the gap in optical gas emission, we also find high gas velocity dispersion breaking into the circumnuclear ring. \citet{CLB2023} found that this part of the ring is where a bar-driven gas inflow connects with the central ring. They found high molecular gas velocity dispersion in this node, similar to what we find. This increased velocity dispersion, seen in both molecular and ionized gas, could disrupt star formation, which would in turn decrease ionized gas emissions. This would explain why this portion of the ring is not marked as an HII region according to the BPT diagram. We also find that the gas velocity dispersion on the opposite (right) side of the ring, where \citet{CLB2023} found another node, is somewhat increased when compared to the rest of the circumnuclear ring. However, we do not see a matching gap in the ring emissions on the right side, so there may be more complicated effects at play.

\subsection{Fueling the central ring and AGN}
Evidence from our models of the gas dynamics, and previous research by \citet{CLB2023}, indicates that the central ring is likely fueled by gas inflows driven by the bar. 

We find that the dynamics of this galaxy are dominated by a rotating disk and a bar. When we fit the line-of-sight velocity maps with a simple disk model, the residuals resembled those found by \citet{LopezCoba22}. They found an S-shaped zero-velocity isophote and antisymmetric residuals at the center in multiple barred galaxies; and, we see both of those features in NGC 5806 (see figure \ref{fig:gasvel_sshape}). Furthermore, the bar model  was the best fit for the velocity fields of this galaxy. We find no evidence of other phenomena or structures significantly affecting the dynamics of this galaxy, aside from potentially the spiral arm. Therefore, we find that the bar of this galaxy has a significant impact on its dynamics. 

Previous studies have found that galactic bars cause galactic inflows towards the galactic center \citep{Sormani2015}. Therefore, we conclude that in NGC 5806, a galaxy that we find to be dynamically dominated by the movement of a galactic bar, there is a bar-driven inflow towards the center of the galaxy. This inflow likely provides fuel for the star formation in the central ring.

Beyond the dynamics of the bar, we observe several other pieces of evidence supporting the existence of a bar-driven gas inflow. As previously discussed, both our observations and \citet{CLB2023} find increased gas velocity dispersion in the north-east and south-west portions of the central ring. \citet{CLB2023} found evidence that these two areas of high velocity dispersion are caused by gas inflows driven by the bar into the central ring, with velocity $V_{in} \approx 120$ km/s and a total mass inflow rate of $5 M_{\odot}$ yr$^{-1}$. 

It is very difficult to estimate the AGN accretion rate.  One common method is to infer the accretion-produced bolometric luminosity, which has been attempted by multiplying the hard X-ray luminosity by an approximate correction factor of ten. According to \citep{Connolly2016}, the 2--10 keV luminosity is $8 \times 10^{39}$ erg/sec, and this observation by SWIFT is consistent with the average of several XMM observations from the 4XMM-DR13 stacked catalog (xsa.esac.esa.int/nxsa-web).
Multiplying the hard X-ray luminosity by a factor of ten gives a bolometric luminosity of $8 \times 10^{39}$ erg/sec, which for an accretion efficiency of 0.1, corresponds to an extremely low mass consumption rate of $10^{-6}$ solar masses/year. 
The Hard X-ray bolometric correction assumed by \citep{Spinoglio2024} would give an accretion rate several times higher. However, if we were instead to scale from the $12\micron$ flux detected by WISE (Band 3, assuming that the active nucleus dominates the total), then the \citep{Spinoglio2024} relations would lead to an accretion rate more than an order of magnitude larger. 

Using the H$\alpha$ flux, we calculate the star formation rate (SFR) in the center 1 x 1 kpc$^2$ of the galaxy. We use the calibration from \citet{Kennicutt1998}: 
\begin{equation}
    SFR(M_{\odot} yr^{-1}) = 7.9 \times 10^{-42}L(H\alpha)(erg s^{-1})
\end{equation}
This calibration does not account for the presence of the AGN, and is often limited in circumnuclear regions due to dust attenuation \citep{Kennicutt1998}. Therefore, we include this calculation only as an estimate and lower boundary on the star formation rate. In the nucleus and central ring of this galaxy, we find that the total star formation rate is $0.06 M_{\odot}$yr$^{-1}$. This SFR falls within the range of SFRs found by \citet{Mazzuca2008} in their study of star-forming rings. 

The gas metallicity of the galaxy provides further evidence for the inflow of gas to the central ring. We find that the gas metallicity of the spiral arm matches that of the central ring, more closely than it matches the gas metallicity in the intermediate region. This suggests that gas from the spiral arm flows inwards to fuel the central ring. If this process occurs very rapidly, it would prevent significant metal enrichment in the intermediate regions prior to the accumulation of gas in the nuclear ring, as suggested by \cite{Bittner2020}. 

All of these pieces of evidence indicate that gas is driven by the bar from the spiral arm into the central ring, fueling the star formation in the central ring. 

Furthermore, in previous research, it has been found that bar-driven inflows can create a gas reservoir in the center of a galaxy \citep{Silva-Lima2022}. This suggests that the gas inflow driven by the bar could be the source of the gas and dust reservoir that we see in the nuclear ring (in optical and IR), which in turn may be an important fuel source for the AGN. It is also possible that the central ring may fuel the AGN in other ways, such as the ring collapse scenario described by \citet{Wada2004}. However, in this galaxy we currently see no explicit movement of gas between the central ring and the AGN. For a full understanding of the feeding of this AGN, especially in regards to the movement of gas in the central region, further study at higher spatial resolution is required.

As described in the section \ref{sec:Introduction}, circumnuclear star-forming rings are often found in barred spiral galaxies \citep{Mazzuca2008}. They are thought to form due to inflowing gas, driven by the bar, being caught in an inner Lindblad resonance. Due to the evidence we have found supporting the dynamics of this galaxy being dominated by bar-driven gas inflows, this seems to be the most likely origin of the central ring that we see in NGC 5806. 

\subsection{Average stellar population}
The central ring has been established as a location of star formation; however, when we calculate the average stellar population, we find surprising results. In the central ring and spiral arm, both areas dominated by star formation, the average stellar age is relatively old and the stellar metallicity seems to be relatively low; comparatively, in the AGN-dominated central and intermediate regions, we tentatively find that the stellar age is younger and the stellar metallicity is higher. This runs counter to what would be expected in regions of star formation, and findings in similar active galaxies, e.g. \citet{Costa-Souza2024}.

These results could indicate that the structure of this galaxy has been present for a long time, and thus the circumnuclear ring and spiral arm have had time to create older stars. It could also suggest that these regions saw higher rates of star formation in the past. One possible explanation for the higher stellar metallicity in the intermediate region is the slightly elevated gas metallicity in this area. Therefore, the counter-intuitive stellar population could be linked to the movement of gas within this galaxy, and thus the feeding and feedback of the AGN.

\section{Conclusions}
\label{SectionConclusion}
We have mapped the gas emission, dynamics, emission line ratios, and average stellar population properties in the central 8 x 8 kpc$^2$ of the active galaxy NGC 5806. 

We confirm that NGC 5806 has an active galactic nucleus,
which is photo-ionizing gas over a wide range of galactic radii. The central region of this galaxy, with a projected radius of about 400 pc, is shown in a BPT diagram to be dominated by an AGN, and shows strong, extended [OIII] emission. In this region we also find a high gas velocity dispersion, up to 90 km/s, and gas metallicity close to solar metallicity. 

Outside of this region, NGC 5806 has a central ring with strong ionized gas emission, which we identify as a star forming region using the BPT diagram. Despite the presence of star formation, we tentatively find that the average stellar population of this ring is relatively old and low metallicity when compared to the stellar population of the rest of the galaxy. The gas metallicity in this ring matches that of the outer spiral arm, supporting a scenario of gas inflow from the spiral arm to the central ring. This inflow is also evidenced by the high gas velocity dispersion in the north-east portion of this ring. When viewing emission line fluxes, the central ring appears to have a gap in the north-east. It is likely that this gap is caused by a dust lane blocking the observer's view of the ring, but star formation may also be reduced in this location due to the inflow of gas from the bar. 

In the intermediate region, between the central ring and the spiral arm, we have found surprisingly far-reaching AGN-dominated gas excitation. We find AGN-dominated gas emissions throughout the intermediate region, extending up to $\sim$3.5 kpc away from the center of the galaxy. We have also shown that this region has slightly higher gas and stellar metallicity than the central ring and the spiral arm. The velocity dispersion of both gas and stars is elevated just outside the central ring.

At the edges of our image, up to 4 kpc away from the AGN, we find an HII region in NGC 5806's spiral arm. This arm's kinematics match the disk around it, and it is has strong, clumpy ionized gas emission. The gas metallicity of this spiral arm is near solar values, and matches the gas metallicity of the central ring.

When looking at the galaxy overall, we have shown that the dynamics are dominated by the presence of a galactic bar. The bar drives gas from the spiral arm into the circumnuclear ring, fueling star formation. We find no evidence of AGN-driven gas outflows, despite the far-reaching AGN-dominated gas excitation. We also see, in the macroscopic view of this galaxy, that the gas is mostly concentrated in the spiral arm and the central ring.


We chose to analyze NGC 5806 in order to observe how an AGN interacts with its host galaxy. In answer to this question, we find AGN-dominated gas excitation in the center of the galaxy, and also reaching over 3 kpc away from the AGN. We also find high stellar and gas velocity dispersion in the very nucleus of the galaxy, potentially driven by the AGN but unresolved in our data. We observe that despite how far the influence of the AGN extends, AGN feedback does not prevent star formation in the central ring. We find unexpected results regarding the average age and metallicity of stars, which may or may not be affected by the AGN. Lastly, we find ways in which the galaxy may affect the AGN: the dynamics of this galaxy are dominated by the bar, rather than the AGN, and this bar drives gas towards the center of the galaxy, creating a circumnuclear ring which may feed the AGN.

\section{acknowledgements}
SIR was supported by the Science and Technology Facilities Council (STFC) of the UK Research and Innovation via grant reference ST/Y002644/1 and by the European Union's Horizon 2020 research and innovation programme under the Marie Sklodowska-Curie grant agreement No 891744. This research has made use of the NASA/IPAC Extragalactic Database (NED), which is operated by the Jet Propulsion Laboratory, California Institute of Technology, under contract with the National Aeronautics and Space Administration.

\bibliography{cite}
\bibliographystyle{aasjournal}
\end{document}